\documentclass{elsart}

\usepackage{bibspacing}
\usepackage{natbib}
\usepackage{times}
\usepackage{xspace}
\usepackage{amssymb}
\usepackage{url}
\usepackage{algorithm,algorithmic}
\usepackage{stmaryrd}
\usepackage{epsfig}
\usepackage[a4paper,margin=1in]{geometry}

\def\prev{\mathbf{X}^{-1}\xspace}

\def\since{\,\mathbf{S}\,}
\def\always{\mathbf{G}^{-1}\,}
\def\sometime{\mathbf{F}^{-1}\,}
\def\impl{\rightarrow}
\def\countq{\mathbf{N}}
\newcommand{\Gp}{\mathbf{G}^{-1}\xspace}

\newcommand{\Fp}{\mathbf{F}^{-1}\xspace}

\newcommand{\ourLTL}{\ensuremath{PTLTL^{FO}}\xspace}
\newcommand{\ourLTLg}{\ensuremath{PTLTL^{FO+}}\xspace}

\newcommand{\unsat}{\textbf{f}}
\newcommand{\sat}{\textbf{t}}

\newcommand{\ds}{\displaystyle\strut}
\newcommand{\urule}[3]{
           \mbox{#1 $\frac{\ds \mbox{#2}}{\ds \mbox{#3}}$}}
\newcommand{\brule}[4]{
            \mbox{#1 $\frac{\ds \mbox{#2}\ \ \ \ \mbox{#3}}
                           {\ds \mbox{#4}}
                     $}
}
\newcommand{\mynode}[2]{\langle #1 \rangle \Downarrow #2}

\newcommand{\idrule}
  {\urule{$(id)$}
         {if $p(\vec t)\!\!\downarrow\; \in h_i$ 
          then $v := \sat$ else $v := \unsat$}
         {$\mynode{h, i, p(\;\vec{t}\;)}{v}$}
}

\newcommand{\relationrule}
  {\urule{$(R)$}
         {if $R(\vec t)\!\!\downarrow$  is true
          then $v := \sat$ else $v := \unsat$}
         {$\mynode{h, i, R(\vec t)}{v}$}
}

\newcommand{\negrule}{
\urule{$(\neg)$}
      {$\mynode{h, i, \psi}{v}$}
      {$\mynode{h, i, \neg \psi}{\neg v}$}
}

\newcommand{\andrule}{
 \brule{$(\land)$}
       {$\mynode{h, i, \psi_1}
                {v_1}$}
       {$\mynode{h, i, \psi_2}{v_2}$}
       {$\mynode
                {h, i, \psi_1 \land \psi_2 }
                {v_1 \land v_2}$}
}

\newcommand{\forallrule}{
 \urule{$(\forall)$}
       {$\mynode
                {h, i, \varphi(\vec t_1)}
                {v_1}
         \qquad \cdots \qquad
         \mynode
                {h, i,\varphi(\vec t_n)}
                {v_n}$}
       {$\mynode
                {h, i, \forall \vec x: p.\varphi(\vec x)}
                {\bigwedge_{i=1}^{n} v_i}$}
        \\ $\qquad$ where 
           $\{\varphi(\vec t_1), \cdots , \varphi(\vec t_n) \}
             = \{\varphi(\vec x) \mid p(\vec x) \in h_i\}$
}

\newcommand{\sincerule}{
 \urule{$(\since)$}
       {$\mynode
                {h,i, \psi_1}
                {v_1}
         \qquad
         \mynode{h,i,\psi_2}{v_2}
         \qquad
         \mynode{h,i-1, \psi_1 \since \psi_2}{v_3}
        $}
       {$\mynode
                {h, i, \psi_1 \since \psi_2}
                {v_2 \lor (v_1\land v_3)}$}
        $i > 1$
}

\newcommand{\sincespecialrule}{
 \urule{$(\since_1)$}
       {$\mynode{h, 1, \psi_2}{v}$}
       {$\mynode
                {h, 1, \psi_1 \since \psi_2}
                {v}$}
}

\newcommand{\prevrule}{
 \urule{$(\prev)$}
       {$\mynode
                {h, i-1, \varphi}
                {v}$}
       {$\mynode
                {h, i, \prev \varphi}
                {v}$}
        $i > 1$
}

\newcommand{\prevspecialrule}{
 \urule{$(\prev_1)$}
       {$v := \unsat$}
       {$\mynode
                {h, 1, \prev \varphi}
                {v}$}
}

\newcommand{\text}[1]{\hbox{#1}}

\journal{ASIACCS'09}

\begin{document}

\setlength{\abovedisplayskip}{7pt}%
\setlength{\belowdisplayskip}{8pt}%

\begin{frontmatter}

  \title{\vspace{-1em}A decidable policy language for\\history-based
    transaction monitoring\vspace{-1em}}

\author{Andreas Bauer, Rajeev Gor\'e, and Alwen Tiu}
\address{Computer Sciences Laboratory, The Australian National University}

\begin{abstract}
  Online trading invariably involves dealings between strangers, so it
  is important for one party to be able to judge objectively the
  trustworthiness of the other. In such a setting, the decision to
  trust a user may sensibly be based on that user's past behaviour. We
  introduce a specification language based on linear temporal logic
  for expressing a {\em policy} for categorising the behaviour
  patterns of a user depending on its transaction history. We also
  present an algorithm for checking whether the transaction history
  obeys the stated policy. To be useful in a real setting, such a
  language should allow one to express realistic policies which may
  involve parameter quantification and quantitative or statistical
  patterns. We introduce several extensions of linear temporal logic
  to cater for such needs: a restricted form of universal and
  existential quantification; arbitrary computable functions and
  relations in the term language; and a ``counting'' quantifier for
  counting how many times a formula holds in the past.  
  We then show that model checking a transaction history against a
  policy, which we call the history-based transaction monitoring
  problem, is \textsc{PSPACE}-complete in the size of the policy formula
  and the length of the history.  The problem becomes decidable in
  polynomial time when the policies are fixed.
  We also consider the problem of transaction monitoring in the case
  where not all the parameters of actions are observable.  We formulate
  two such ``partial observability'' monitoring problems, and show their
  decidability under certain restrictions.

\end{abstract}

\end{frontmatter}

\section{Introduction}
\vspace{-0.5em}
Internet mediated trading is now a common way of exchanging goods and
services between parties who may not have engaged in transactions with
each other before.
The decision of a seller/buyer to engage in a transaction is usually
based on the ``reputation'' of the other party, which is often provided
via the online trading system itself.  These so-called \emph{reputation
  systems} can take the form of numerical ratings, which can be computed
based on feedback from users (cf.\ \cite{josang07dss} for a survey of
reputation systems) While many reputation systems used in practice seem
to serve their purposes, they are not without problems (cf.\
\cite{josang07dss}) and can be too simplistic in some cases.  For
example, in eBay.com, the rating of a seller/buyer consists of two
components: the number of positive feedbacks she gets, and the number of
negative feedbacks. A seller with, say 90 positive feedbacks and 1
negative feedback may be considered trustworthy by some.  But one may
want to correlate a feedback with the monetary value of the transaction
by checking if the one negative feedback was for a very expensive item,
or one may want to check other more general relations between different
parameters of past transactions.

Here, we consider an alternative (and complementary) method to describe
the reputation of a seller/buyer, by specifying explicitly what
constitutes a ``good'' and a ``bad'' seller/buyer based on the observed
patterns of past transactions.  More specifically, we introduce a formal
language based on linear temporal logic for encoding the desired
patterns of behaviours, and a mechanism for checking these patterns
against a concrete history of transactions.
The latter is often referred to as the {\em monitoring
  problem} since the behaviour of users is being monitored,
but here, it is just a specific instance of model checking for temporal
logic.
The patterns of behaviours, described in the logical language, serve as
a concise description of the policies for the user on whether to engage
with a particular seller/buyer.  The approach we follow here is
essentially an instance of {\em history-based access control} (see e.g.,
\cite{edjlali98ccs,havelund02tacas,fong04sp,bartoletti05fossacs,krukow05ccs,bauer:leucker:schallhart:FSTTCS06}).
More precisely, our work is closely related to that of Krukow et
al.\ \cite{krukow05ccs,krukow08jcs}.

There are two main ideas underlying the design of our language:

\begin{description}
\item[\rm \em Transactions vs.\ individual actions:] Following Krukow et
  al., we are mainly interested in expressing properties about
  transactions seen as a logically connected grouping of actions, for
  example because they may represent a run of a protocol. A history in
  our setting is a list of such transactions.  This is in contrast to
  the more traditional notion of history as a list of individual actions
  (i.e., a trace), e.g., as in \cite{edjlali98ccs,havelund02tacas},
  which is common in monitoring program execution.

\item[\rm \em Closed world assumption:] The main idea underlying the
  design of our quantified policies is that a policy should only express
  properties of objects which are observed in the history.  For example,
  in monitoring a typical online transaction, it makes sense to talk
  about properties that involve ``all the payments that have been
  made''. Thus, if we consider a formalisation of events using
  predicates, where $pay(100)$ denotes the payment of $100$ dollars
  (say), then we can specify a policy like the one below left which
  states that all payments must obey $\psi$:
  $$
  \forall x.\ pay(x) \impl \psi(x) \qquad \qquad \qquad
  \forall x.\ \neg pay(x) \impl \psi(x)
  $$
  However, it makes less sense to talk about ``for all dollar amounts
  that a seller did not pay'', like the policy above right, since this
  involves infinitely many possibility (e.g., the seller paid 100, but
  did not pay 110, did not pay 111, etc.).  We therefore restrict our
  quantification in policies to have a ``positive guard'',
  guaranteeing that we always quantify over the finitely many values
  that have already been observed in the history.
\end{description}

An important consequence of the closed world assumption is that we can
only describe relations between known individual objects. Thus we can
enrich our logical language with computable functions over these
objects and computable relations between these objects without losing
decidability of the model checking problem.  One such useful extension
is arithmetic, which allows one to describe constraints on various
quantities and values of transactions.

Our base language for describing policies is the pure past fragment of
linear temporal logic \cite{Pnueli77} since it has been used quite
extensively by
others~\cite{roger01csfw,havelund02tacas,krukow05ccs,bauer:leucker:schallhart:FSTTCS06}
for similar purposes. However, the following points distinguish our
work from related work in the literature:
\begin{itemize}
\item We believe our work is the first to incorporate both quantified
  policies and computable functions/relations within the same logic.
  Combining unrestricted quantifiers with arbitrary computable
  functions easily leads to undecidability (see Section~\ref{sec:guard}).
\item We extend temporal logic with a ``counting quantifier'', which
  counts how many times a policy has been satisfied in the past. A
  similar counting mechanism was proposed in
  \cite{krukow05ccs,krukow08jcs} as a part of a meta-policy language.
  But in our work, it is a part of the same logic.

\item We consider new monitoring problems based on a notion of {\em
    partial observability} which seem to arise quite naturally in online
  trading platforms where a user (or a system provider) cannot directly
  observe all parameters of an action. For instance, in eBay, it may not
  be always possible to observe whether payments have been made, or it
  may be possible to observe a payment but not the exact amount paid. We
  model unobservable parameters in an action as variables representing
  unknown values. Given a policy and a history containing unknown
  parameters, we ask whether the policy is satisfied under some
  substitution of the variables (the {\em potential satisfiability
    problem}), or under {\em all} substitutions (the {\em adherence
    problem}).
\end{itemize}

The rest of the paper is organised as follows.
Section~\ref{sec:language} introduces our policy language $\ourLTL$, for
``past time linear temporal logic with first-order (guarded)
quantifiers'', and defines its semantics.
Section~\ref{sec:examples} presents some examples using $\ourLTL$ for
specifying access control policies. Two examples are formalisations of
known security policies, which are trace-based in the sense that the
histories are just traces, and that go beyond the scope of online
trading systems alone.  The third example shows a transaction-based
policy as it can be used for eBay.com type of systems.
Section~\ref{sec:modelchecking} considers the model checking problem for
$\ourLTL$ which we show to be \textsc{pspace}-complete, even if we
restrict it to what we call trace-like histories.
Fixing the policies reduces the
complexity to \textsc{ptime}.  
Section~\ref{sec:count} presents an extension of $\ourLTL$ with a
counting quantifier allowing us to express that a policy depends on
the number of times another policy was satisfied in the past. 
The model checking problem for this extension remains
\textsc{pspace}-complete.  In Section~\ref{sec:partial}, we consider
more general (undecidable) monitoring problems where not all the
parameters of an action can be observed.
By restricting the class of allowed functions and relations,
we can obtain decidability of both the potential satisfiability and
adherence problems,
for example,
when the term
language of the logic is restricted to linear arithmetic.
Section~\ref{sec:guard} discusses possible decidable extensions to the guarded
quantifiers.  
Section~\ref{sec:conc} concludes the paper and discusses related work.
Detailed proofs are given in the Appendix.


\vspace{-1em}

\section{The policy language: definitions and notation}
\vspace{-0.5em}
\label{sec:language}

Since we are interested in the notion of history-based access control,
our definition of history is a simplification of that of
\cite{krukow08jcs}.  A history is organised as a list of sessions. Each
session is a finite set of events, or actions. Each event is represented
by a predicate. A session represents a ``world'' in the sense of a
Kripke semantics where the underlying frame is linear and discrete.

The term structures of our policy language are made up of variables and
interpreted multi-sorted function symbols. Function symbols of zero
arity are called {\em constants}. Terms are ranged over by $s,t,u$.
Variables of the language, denoted by $x,y,z$, range over certain
domains, such as strings, integers, or other finite domains.  We call
these domains {\em base types} or simply {\em types}.  We assume a
distinguished type $prop$ which denotes the set of propositions of the
logic, and which must not be used in the types of the function symbols
and variables.  That is, we do not allow logical formulae to appear at
the term level.  Function symbols and variables are typed.

We assume an interpretation where distinct constants of the same type
map to distinct elements of the type. We shall use the same symbol, say
$a$, to refer both to an element of some type $\tau$ and the constant
representing this element.  Function symbols of one or more arities
admit a fixed interpretation, which can be any total recursive function.
We shall assume the usual function symbols for arithmetic, $+$, $-$,
$\times$, etc., with the standard interpretations. The language we are
about to define is open to additional interpreted function symbols,
e.g., string related operations, etc.  We shall use $f,g,h$ to range
over function symbols of arity one or more, and $a,b,c,d$ to range over
constants.  We also assume a set of interpreted relations, in
particular, those for arithmetic, e.g., $<$, $=$, $\geq$, etc. These
interpreted relations are ranged over by $R$.  All the interpreted
functions and relations have first-order types, i.e., their types are of
the form
$$
\tau_1 \times \cdots \times \tau_n \rightarrow \tau
$$
where $\tau$ and $\tau_1,\ldots,\tau_n$ are base types.
We shall restrict to computable relations
$R$. Of course, there is also the (rigidity) assumption that the
function $f$, constant $c$ and relation $R$ have the same fixed
interpretation over all worlds.

Since our term language contains interpreted symbols, we assume that
there is a procedure for evaluating terms into values. We also assume
that each term can be evaluated to a unique value.  Given a term $t$, we
shall denote with $t\downarrow$ the unique value denoted by this term,
e.g., if $t = (2 + 3)$ then $t\downarrow = 5$.  Given an atomic formula
$p(t_1,\ldots,t_n)$, we shall write $p(t_1,\ldots,t_n)\downarrow$ to
denote $p(t_1\downarrow,\ldots,t_n\downarrow).$ The policy language is
given by the following grammar:
$$
  \psi ::= p(t_1,\ldots, t_m) \mid R(t_1,\ldots,t_n) \mid \psi \land \psi \mid \neg \psi 
\mid \prev \psi \mid \psi \since \psi \mid \forall (x_1,\ldots,x_n):p.\ \psi.
$$ 
In the quantified formula $\forall (x_1,\ldots,x_n):p.\ \psi$, 
where $n \geq 1$, the
symbol $p$ is an $n$-ary predicate of type
$\tau_1 \times \cdots \times \tau_n \rightarrow prop$, and each $x_i$ is
of type $\tau_i$.  The intended interpretation of this quantification is
that the predicate $p$ defines a subtype of $\tau_1 \times \cdots \times
\tau_n$, which is determined by the occurrence of $p$ in the world
(session) in which the formula resides.  For example, in a world
consisting of
$
\{p(1,1), p(1,2), p(1,3), q(4) \} 
$
the predicate $p$ represents the set $\{(1,1), (1,2), (1,3) \}$, i.e., a
subset of $N \times N$.  We shall often abbreviate $\forall
(x_1,\ldots,x_n):p.\ \psi$ as simply $\forall \vec x:p.\ \psi$ when the
exact arity and the information about each $x_i$ is not important or can
be inferred from context. The notions of free and bound variables are
defined as usual. A formula is {\em closed} if it has no occurrences of
free variables.
  \begin{figure}[t]
    \centering
  \begin{eqnarray*}
    (h,i) & \models & p(t_1,\ldots,t_n) \text{ iff } 
    p(t_1\downarrow, \ldots, t_n\downarrow) \in h_i
    \\
    (h,i) & \models & R(t_1,\ldots,t_n) \text{ iff } 
    R(t_1\downarrow, \ldots, t_n\downarrow) \text{ is true }
    \\
    (h,i) & \models &  \psi_1 \land \psi_2 \text{ iff } 
    (h,i) \models \psi_1 \text{ and } (h,i) \models \psi_2
    \\
    (h,i) & \models &  \neg \psi \text{ iff } 
    (h,i) \not \models \psi
    \\
   (h,i) & \models &  \prev \psi \text{ iff } 
    i > 1 \text{ and } (h, i-1) \models \psi
    \\
   (h,i) & \models &  \psi_1 \since \psi_2 \text{ iff } 
   \text{ there exists } j \leq i
    \text{ such that } (h,j) \models \psi_2 \text{ and } 
    \\     &         & \text{ for all } k, \text{ if } j < k \leq i
    \text{ then } (h,k) \models \psi_1
    \\
   (h,i) & \models &  \forall (x_1,\ldots,x_n):p.\ \psi \text{ iff }
   \text{ for all }
    c_1,\ldots,c_n, \text{ if } p(c_1,\ldots,c_n) \in h_i  \\
         &         & \text{ then } (h,i) \models \psi[x_1 := c_1, \ldots, x_n := c_n].
  \end{eqnarray*}
    \caption{Semantics of $\ourLTL$}
    \label{fig:semantics}
  \end{figure}
\begin{defn}
  An {\em event} (or an {\em action}) is a predicate $p(c_1,\ldots,c_n)$
  where each $c_i$ is a constant and $p$ is an uninterpreted predicate
  symbol. A {\em session} is a finite set of events.  A {\em history} is
  a finite list of sessions.
\end{defn}
A standard definition for the semantics of first-order logic uses a
mapping of free variables in a formula to elements of the types of the
variables.  To simplify the semantics, we shall consider only closed
formulae.  The semantics for quantified statements is then defined by
closing these statements under variable mappings. We use the
notation $\sigma$ and $\theta$ to range over partial maps from variables
to elements of types.  We usually enumerate them as, e.g.,
$[x_1 := a_1,\ldots,x_n := a_n].$ Since we identify a constant with the
element represented by that constant, a variable mapping is both a
semantic and a syntactic concept. The latter means that we can view a
variable mapping as a substitution. Given a formula $\psi$ and variable
mapping $\sigma$, we write $\psi \sigma$ to denote a formula resulting
from replacing each free variable $x$ in $\psi$ with the constant
$\sigma(x)$.  From now on, we shall use the term variable mapping and
substitution interchangeably.

The semantic judgement that we are interested in is of the form $(h,i)
\models \psi$, where $h$ is a history, $i$ is an index referring to the
$i$-th session in $h$, and $\psi$ is a closed formula. The judgement
reads ``$\psi$ is true at the $i$-th world in the history $h$''.  We denote
with $|h|$ the length of $h$, and with $h_i$ the $i$-th element of $h$
when $i \leq |h|$.
\begin{defn}
  The {\em forcing relation} $(h,i) \models \psi$, where $h$ is a
  history, $i$ an integer, and $\psi$ a formula, is defined inductively
  as shown in Figure~\ref{fig:semantics} where $1 \leq i \leq |h|$.
  We denote with $h \models \psi$ the relation $(h, |h|) \models \psi.$
  The boolean connectives $\lor$ (disjunction) and $\impl$ (implication)
  are defined in the standard way using negation and conjunction.
  We derive the operators $\Fp \varphi \equiv \top \since \varphi$ (``sometime in the past''), 
  and $\Gp \varphi \equiv \neg \Fp (\neg \varphi)$ (``always in the past''), 
 where $\top$ (``true'') is short for $p \vee \neg p$.
\end{defn}

Note that allowing unrestricted quantifiers can cause model checking 
to become undecidable, depending on the interpreted functions and relations.
For example, if we allow arbitrary arithmetic expressions in the term
language, then we can express solvability of Diophantine equations, which is undecidable~\cite[Chapter~5]{matiyasevich93book}.


\vspace{-1em}

\section{Some example policies}
\vspace{-0.5em}
\label{sec:examples}

Let us now examine some example policies known from the literature, and
our means of expressing them concisely and accurately.  We also examine
some policies from applications other than monitoring
users in online trading systems to demonstrate that our language can
model the requirements of other related domains as well if they
can be expressed as trace-based properties.

\textbf{One-out-of-k policy.}
The \emph{one-out-of-k policy} as described in \cite{edjlali98ccs} concerns the
monitoring of web-based applications.
More specifically, 
it concerns monitoring three specific situations: connection to a remote site, opening local files,
and creating subprocesses.  We model this as follows, with 
the set of events being
\begin{description}
\item[\rm $open(file,mode)$:] request to open the file $file$ in mode,
  $mode$, where $file$ is a string containing the absolute path, and
  $mode$ can be either $\mathrm{ro}$ (for read-only) or $\mathrm{rw}$
  (for read-write).  There can be other modes but for simplicity
  we assume just these two;
\item[\rm $read/write/create(file)$:] request to read/write/create a file;
\item[\rm $connect$:] request to open a socket (to a site which is
  irrelevant for now);
\item[\rm $subproc$:] request to create a subprocess. 
\end{description}
We assume some operators for string manipulation: the function $path(file)$
which returns the absolute path to the directory in which the file resides,
and the equality predicate $=$ on strings. 
The history in this setting is restricted to one in which every
session is a singleton set. 
We now show how to encode one of the policies as described in \cite{edjlali98ccs}:
allow a program to open local files in user-specified
directories for modifications if and only if it has created them, and it
has neither tried to connect to a remote site nor tried to create a
sub-process.
Suppose that we allow
only one user-specified directory called ``Document''. 
Then this policy can be expressed as:
$$
\begin{array}{ll}
\forall (x,m) : open. 
m = \mathrm{rw} \impl [ & path(x) = \hbox{``Document''} 
\land ~ \sometime create(x) ~\land \\
 & \neg \sometime connect ~ 
 \land ~ \neg \sometime subproc].
\end{array}
$$

\textbf{Chinese wall policy.}
The chinese wall policy \cite{brewer89} is a common access control
policy used in financial markets for managing conflicts of interests.
In this setting, each object for which access is requested, is
classified as belonging to a {\em company dataset}, which in turn
belongs to a {\em conflict of interest class}.  The idea is that a user
(or subject) that accessed an object that belonged to a company $A$ in
the past will not be allowed to access another object that belongs to a
company $B$ which is in the same conflict of interest class as $A$.  

To model this policy, we assume the following finite sets: $U$ for
users, $O$ for objects, $D$ for company datasets, and $C$ for the
names of the conflict of interest class.  The event we shall be
concerned with is access of an object $o$ by a user $u$. We shall
assume that this event carries information about the company
dataset to which the object belongs, and the name of the conflict
of interest class to which the company dataset belongs.  That is,
$access$ is of type $U \times O \times D \times C \rightarrow prop$.
A history in this case is a sequence of singleton sets containing the
$access$ event.  The policy, as given in \cite{brewer89}, specifies
among others that
\begin{quote}
``access is only granted if the object requested:
\begin{enumerate}
\item is in the same company dataset as an object already accessed
by that subject, or
\item belongs to an entirely different conflict of interest class.''
\end{enumerate}
\end{quote}
Implicit in this description is that first access (i.e., no prior history) 
is always allowed. We can model the case where no prior history exists
simply using the formula $\neg \prev \top.$
This policy can be expressed in our language as follows:
$$
\begin{array}{l}
\forall (s,u,d,c) : access.\ \neg \prev \top ~ \lor \\ 
\qquad (\prev \sometime \exists (s',u',d',c'): access.\ s = s' \land d = d') ~ 
 \lor \\
\qquad (\prev \always \forall (s',u',d',c'): access.\ s = s' \impl \neg (c = c')).
\end{array}
$$

\textbf{eBay.com.}
In this example, we consider a scenario where a potential buyer wants to
engage in a bidding process on an online trading system like eBay.com,
but the buyer wants to impose some criteria on what kind of sellers she
trusts.  A simple policy would be something like ``only deal with a
seller who was never late in delivery of items''.  In this model, a
session in a history represents a complete exchange between buyer and
seller, e.g., the bidding process, winning the bid, payment,
confirmation of payment, delivery of items, confirmation of delivery, and
the feedbacks.  We consider the following events (we are considering the
history of a seller):
\begin{description}
\item[\rm  $win(X,V)$:] the bidder won the bid for item $X$ for value $V.$
\item[\rm  $pay(T,X,V)$:] payment of the item $X$ at date $T$ of the sum $V$
  (numerical value of dollars).
\item[\rm  $post(X,T)$:] the item $X$ is delivered within $T$ days.
  \footnote{Note that in the actual eBay system, no concrete number of
    days is given, but instead buyers can rate the time for posting and
    handling in the feedback forums in a range of 1 to 5.}
\item[\rm $negative$, $neutral$, $positive$:] represents, respectively, negative,
neutral and positive feedbacks.
\end{description}
There are of course other actions and parameters that we can formalise,
but these are sufficient for an illustration.  Now, suppose the buyer
sets a criterion such that a posting delay greater than 10 days
after payment is unacceptable.  This can be expressed simply as:
\begin{equation}
\label{eq:ontime}
\always [
\forall (t,x,v) : pay.\ \exists (y,t') : post.\ x = y \land t' \leq 10].
\end{equation}
Of course, for such a simple purpose, one can rely on eBay's rating
system, which basically computes the number of feedbacks in each
category (positive, neutral and negative). However, the seller's rating
may sometimes be too coarse a description of a seller's reputation. For
instance, one is probably willing to trust a seller with some negative
feedbacks, as long as those feedbacks refer to transactions involving
only small values.  A buyer can specify that she would trust a seller
who never received negative feedbacks for transactions above a certain
value, say, 200 dollars.  This can be specified as follows:
$
\always [\forall (t,x,v) : pay.\ v \geq 200 \impl \neg negative].
$


\vspace{-1em}

\section{Model checking \ourLTL}
\vspace{-0.5em}
\label{sec:modelchecking}

\begin{figure}[t]
\centering
\begin{tabular}[c]{ccc}
\idrule
$\qquad$
\relationrule
\\ \\
\negrule
$\qquad$
\andrule
\\ \\
\forallrule
\\ \\
\sincerule
\\ \\
\sincespecialrule
$\quad$
\prevrule
$\quad$
\prevspecialrule
\\
\end{tabular}
  \caption{Evaluation rules for deciding whether $(h,i) \models \varphi$.}
  \label{fig:myeval}
\end{figure}

Let us now consider the model checking problem for $\ourLTL$, i.e.,
deciding whether $h \models \varphi$ holds. 
We shall see that the model checking problem is \textsc{pspace}-complete,
even in the purely logical case, i.e., the case where no
interpreted functions or relations occur in the formula. 

We prove the complexity of our model checking problem via a
terminating recursive algorithm.  The algorithm is presented
abstractly via a set of rules which successively transform a triple
$\langle h, i, \varphi \rangle$ of a history, an index and a formula,
and return a truth value of either $\sat$ or $\unsat$ to indicate that
$(h,i) \models \varphi$ (resp.\ $(h,i) \not \models \varphi$).
We  write $\mynode {h, i, \varphi}{v} $
to denote this relation and
overload the logical connectives $\land$, $\lor$ and $\neg$
to denote operations on boolean values, e.g., $\sat \land \sat = \sat$, etc.
Since $\psi_1 \since \psi_2 \equiv \psi_2 \lor (\psi_1 \land \prev (\psi_1 \since \psi_2 ))$,
we shall use the following semantic clause for 
$\psi_1 \since \psi_2$ which is equivalent to the original one:
$$
(h,i) \models \psi_1 \since \psi_2 \text{ iff } (h,i) \models \psi_2
\text{ or } [(h,i) \models \psi_1 \text{ and } i > 1 \text{ and }
(h,i-1) \models \psi_1 \since \psi_2].
$$
The rules for the evaluation judgement are given in
Figure~\ref{fig:myeval}.  
To evaluate the truth value of
$\langle h, i, \varphi \rangle$, we start with the judgement
$\mynode{h,i,\varphi}{v}$ where $v$ is still unknown.  We then
successively apply the transformation rules bottom up, according to
the main connective of $\varphi$ and the index $i$.  Each
transformation step will create $n$-child nodes with $n$ unknown
values.  Only at the base case (i.e., $id$, $R$, or $\prev_1$) the
value of $v$ is explicitly computed and passed back to the parent
nodes.  A run of this algorithm can be presented as a tree whose nodes
are the evaluation judgements which are related by the transformation
rules.  
A straightforward simultaneous induction on the derivation tree of the
evaluation judgements
yields:

\begin{lem}
\label{lm:eval}
The judgement $\mynode{h, i, \varphi}{\sat}$ is derivable if and only if
$(h,i) \models \varphi$ and the judgement $\mynode{ h, i, \varphi}{\unsat}$
is derivable if and only if $(h,i) \not \models \varphi.$
\end{lem}

\begin{thm}
\label{thm:modelchecking}
  Let $\varphi$ be a $\ourLTL$ formula and $h$ a history.  
  If the interpreted functions and relations in $\varphi$ are in \textsc{pspace}, then 
  deciding 
  whether $h \models \varphi$
  holds is \textsc{pspace}-complete. 
\end{thm}

Although the model checking problem is \textsc{pspace}-complete, in
practice, one often has a fixed policy formula which is evaluated
against different histories.  Then, it makes sense to ask about the
complexity of the model checking problem with respect to the size of
histories only (while restricting ourselves to interpreted functions
and relations computable in polynomial time).
\begin{thm}
  \label{thm:fixed}
  The decision problem for $h \models \varphi$, where $\varphi$ is
  fixed, is solvable in polynomial time. 
\end{thm}
An easy explanation for the above hardness result is via a
polynomial time encoding of the \textsc{PSPACE}-complete QBF-problem
(cf.\ \cite{524279} and Appendix).
Given a boolean expression like
$E(x_1,x_2, x_3) \equiv (x_1 \vee \lnot {x_2}) \wedge (\lnot {x_2} \vee x_3)$
and the QBF-formula
$
F \equiv \forall x_1.\ \exists x_2.\ \forall x_3.\ E(x_1,x_2,x_3)
$,
we can construct a corresponding \ourLTL-formula, $\varphi \equiv
\forall x_1:p_1.\ \exists x_2:p_2.\ \forall x_3:p_3.\ E'(x_1, x_2, x_3)$
where 
$E'(x_1, x_2, x_3) \equiv (true(x_1) \vee
\neg true(x_2)) \wedge (\neg true(x_2) \vee true(x_3))$, and a history,
$h$ below, representing all possible interpretations of $F$'s variables in
a single session:
$$
h = \{ p_1(0), p_1(1), p_2(0), p_2(1), p_3(0),p_3(1),
true(1) \}.
$$
It is then easy to see that $F$ evaluates to $\top$ if and only if $h
\models \varphi$ holds. Thus solving our general model checking
problem, like QBF, may require time exponential in the number of
quantifiers.

On the surface it seems that this ``blow up'' is caused by the multiple
occurrences of the same predicate symbol in a single session.
It is therefore natural to ask whether the complexity of the problem can
be reduced if we consider histories where every predicate symbol can
occur at most once in every session.
Surprisingly, however, even with this restriction, model checking
remains \textsc{pspace}-complete.  
Consider, for example, the following polynomial encoding of the above
QBF-instance, using this restriction:
$$
\begin{array}{l}
\{p_3(0), true(1) \};\{p_3(1), true(1) \}; \ldots;
\{p_1(0), true(1)\} ; \{p_1(1), true(1)\}
\models \\
\qquad 
\always \forall x_1 : p_1.\ 
\Fp \exists x_2 : p_2.\ 
\always \forall x_3 : p_3.\ E'(x_1, x_2, x_3)).
\end{array}
$$

\begin{defn}
A history $h$ is said to be {\em trace-like} if for all $i$ such that
$1 \leq i \leq |h|$, for all $p$, ${\vec t}$ and ${\vec s}$, if $p({\vec t}) \in h_i$
and $p({\vec s}) \in h_i$, then ${\vec t} = {\vec s}.$
\end{defn}
\begin{thm}
  \label{thm:tracelike}
  Let $\varphi$ be a $\ourLTL$ formula and $h$ a trace-like history.
  If the interpreted functions and relations in $\varphi$ are in \textsc{pspace},
  then 
 deciding whether 
 $h \models \varphi$ holds is
  \textsc{pspace}-complete.
\end{thm}


\vspace{-1em}

\paragraph*{Implementation.}
\vspace{-0.5em}
\label{sec:impl}

We have implemented the above in terms of a prototypic model checker for
\ourLTL, which can be freely downloaded 
 and evaluated at
\textsf{http://code.google.com/p/ptltl-mc/}.
The model checker primarily accepts two user inputs: a \ourLTL policy
and a history which is then checked against the policy. We use
FOL-RuleML \cite{boley2004frf} as the input format for the policy
since it is due for standardisation as the W3C's first-order logic
extension to RuleML \cite{ruleml}.
Thus users can even specify policies using
graphical XML-editors with a FOL-RuleML DTD extended by our temporal
operators.

Our model checker is currently not optimised for performance, but
it demonstrates the feasibility and practicality of our approach to
tackling these problems, as its main algorithm is based directly on the
rules from Figure~\ref{fig:myeval}.  
The above web site contains Ocaml source code (as well as a
statically linked binary for Linux) and some example policies from
Section~\ref{sec:examples} in XML-format.


\vspace{-1em}

\section{Extending \ourLTL with a counting quantifier}
\vspace{-0.5em}
\label{sec:count}

We now consider an extension of our policy language with a counting
quantifier. The idea is that we want to count how many times a policy
was satisfied in the past, and use this number to write another
policy.  

The language of formulae is extended with the 
construct 
$
\countq x : \psi.\ \phi(x)
$
where $x$ binds over the formula $\phi(x)$ and is not free in $\psi.$
The semantics of this formula is as follows:
\begin{quote}
$(h, i) \models \countq x : \psi.\ \phi(x)$ iff $(h,i) \models \phi(n),$
where 
$n = |\{j \mid \hbox{$1 \leq j \leq i$ and $(h,j) \models \psi$} \}|.$
\end{quote}

Krukow et al. also consider a counting operator, $\#$, which applies to a
formula. Intuitively, $\# \psi$ counts the number of sessions in which
$\psi$ is true, and 
can be used inside other arithmetic
expressions like $\# \psi \leq 5$. The advantage of our approach is
that we can still maintain a total separation of these arithmetic
expressions and other underlying computable functions from the logic,
thus allowing us to modularly extend these
functions. Another notable difference is that our extension resides in
the logic itself, instead of a separate ``meta'' policy language like
theirs.

{\em Examples: } 
For
example, 
we show how to state a ``meta'' policy
such as:
``engage only with a seller whose past transactions with
negative feedbacks constitute at most a quarter of the total
transactions''. 
This can be expressed succinctly by the following
formula since $\countq y : \top$ instantiates $y$ to be the length of
the transaction history to date:
$$
\countq x : negative.\ \countq y : \top.\ \frac{x}{y} \leq \frac{1}{4}.
$$
A more elaborate example is
the formula in
Equation~\ref{eq:ontime} without the $\always$-operator:
$$
\psi \equiv 
\forall (t,x,v) : pay.\ \exists (y,t') : post.\ x = y \land t' \leq 10.
$$
Then one can specify a policy that demands that ``the seller's delivery is
{\em mostly} on-time'', where {\em mostly} can be given as a
percentage, such as $90\%$, via:
$$
\countq x : \psi.\ \countq y : \top.\ \frac{x}{y} \leq 0.9.
$$

The proof of the theorem below is a straightforward extension of
the proof of Theorem~\ref{thm:modelchecking}.

\begin{thm}
\label{thm:counting}
Assuming that the interpreted functions and relations are in \textsc{pspace},
the model checking problem for $\ourLTL$ extended with the counting 
quantifier is \textsc{pspace}-complete.
\end{thm}


\vspace{-1em}

\section{Partial observability}
\vspace{-0.5em}
\label{sec:partial}

In some online transaction systems, like eBay, certain events may not
be wholly observable all the time, even to the system providers, e.g.,
payments made through a third-party outside the control
of the provider.  
\footnote{
   eBay asks the user for confirmation of payment, but does not check
   whether the payment goes through. In our simplified account, this is
   modelled by an unknown amount in the payment parameters.  } 
We consider scenarios where some information is missing from the history
of a client (buyer or seller) and the problem of enforcing security
policies in this setting.%

{\em Examples:} Consider the policy $ \psi \equiv \always [\forall
(x,v) : win. \exists (t,y,u) : pay.  x = y \land v = u] $ which states
that every winning bid must be paid with the agreed dollar amount.
The history below,
where $X$ represents an unknown amount, can
{\em potentially satisfy} $\psi$ when $X = 100$ (say): 
$$
\begin{array}{ll}
h = & \{win(a,100), pay(1,a,100), post(a,5)\} ; \\
  & \{win(a,100), pay(2,a,X),post(a,4), positive\}
\end{array}
$$
Of course it is also possible that
the actual amount paid is less than 100, in which case
the policy is not satisfied.
There are also cases in which the values of the unknowns
do not matter.
For instance, a system provider may not be able to verify payments, but it 
may deduce
that if a buyer leaves a positive
remark, that payment has been made. 
That is, a policy like the following:
$$
\varphi' \equiv \always [\forall (x,v) : win.
\exists (t,y,u) : pay. x = y \land (u = v \lor positive)]
$$
which checks that a payment was made and it was made for exactly the same amount
as the winning bid, or the transaction is concluded with a positive
feedback (which presumably means everything is fine).  In this case,
we see that $h$ still satisfies $\varphi'$ under all substitutions for
$X$.

We consider two problems arising from partial observability.
For this, we extend slightly the notion of history and sessions.

\begin{defn}
A {\em partially observable session}, or {\em po-session} for short,
is a finite set of predicates of the form $p(u_1,\ldots,u_n)$,
where $p$ is an uninterpreted predicate symbol and each $u_i$ is
either a constant or a variable.
A {\em partially observable history} ({\em po-history}) is
a finite list of po-sessions.
Given a po-history $h$, we denote with $V(h)$ the set of variables occurring in $h$.
\end{defn}

\begin{defn}
Given a po-history $h$, a natural number $i$, and a closed formula $\psi$, we say that
{\em $h$ potentially satisfies $\psi$ at $i$}, written $(h,i) \vdash \psi$, if
there exists a substitution $\sigma$ such that $dom(\sigma) = V(h)$ and
$(h\sigma, i) \models \psi.$
We say that {\em $h$ adheres to $\psi$ at $i$}, written $(h,i) \Vdash \psi$,
if $(h\sigma, i) \models \psi$ for all $\sigma$ such that $dom(\sigma) = V(h).$
\end{defn}

Notice that the adherence problem is just the dual of the potential
satisfiability problem. That is, $(h,i) \Vdash \psi$ if and only if
$(h,i) \not \vdash \neg \psi.$ In general the potential satisfiability
problem is undecidable, since one can easily encode solvability of 
general Diophantine equations, which is known to be undecidable. To
see this, let us suppose that the term language of the logic includes
standard arithmetic operators (including exponentiation).  Then we can
express directly any Diophantine equations within our term language.
Let us denote with $D(x_1,\ldots,x_n)$ a set of Diophantine equations
whose variables are among $x_1, \ldots, x_n$. Assume that we have $n$
uninterpreted unary predicate symbols $p_1, \cdots , p_n$ which take an
integer argument. Then solvability of $D(x_1,\ldots,x_n)$ is reducible
to the satisfiability problem
$$
\{p_1(x_1), \ldots, p_n(x_n)\} \vdash \exists x_1 : p_1. \cdots \exists x_n : p_n.\psi(x_1,\ldots,x_n)
$$ 
where $\psi(x_1,\ldots,x_n)$ is the conjunction of all the equations in 
$D(x_1,\ldots,x_n).$
So obviously decidability of the potential satisfiability problem is
dependent on the term language of the logic. We consider here the
decidability problem for the case where the term language is the
language of linear arithmetic over integers, i.e., terms of the form
(modulo associativity and commutativity of $+$):
$
k_1x_1 + \cdots + k_nx_n + c,
$
where $c$ and each $k_i$ are integers. We also assume the standard relations
on integers $=$, $\geq$ and $\leq.$ 
It is useful to introduce a class of {\em constraint formulae}
generated from the following grammar:
$$
C ::= \top \mid \bot \mid t_1 = t_2 \mid t_1 \leq t_2 \mid t_1 \geq t_2 \mid
C_1 \land C_2 \mid C_1 \lor C_2 \mid \neg C.
$$
We say that a constraint $C$ is {\em satisfiable} if there exists
a substitution $\sigma$ such that $C\sigma$ is true. Satisfiability of
constraint formulae is decidable (see \cite{kroening08book} for 
a list of algorithms).
The decidability proof of the potential satisfiability problem 
involves a transformation of the judgement $(h,i) \vdash \psi$  
into an equivalent constraint formula.

\begin{thm}
\label{thm:partial}
  The potential satisfiability problem and the adherence problem for
  $\ourLTL$ with linear arithmetic are decidable.
\end{thm}

We note that the transformation of the potential satisfiability problem
to constraints formulae used in the proof of Theorem~\ref{thm:partial}
may result in an exponential blow-up. 
But if we fix the formula, we may be able to obtain a polynomial 
translation, in the size of the history. We leave the details of this 
and other restrictions to future work.


\vspace{-1em}

\section{Extended guarded quantifiers}
\vspace{-0.5em}
\label{sec:guard}

As we have mentioned in the introduction, an underlying
design principle for our quantified policies is the closed-world 
assumption (CWA). The guarded quantifier in $\ourLTL$ is
the most basic quantifier, and by no means the only one 
that enforces this CWA principle.
It is a natural theoretical question 
to ask what other possible extensions achieve the same effect,
although we have not so far seen the need for them in practice. 

We have mentioned earlier that introducing negation in the guard 
easily leads to undecidability. Surprisingly, simple extensions
with unrestricted disjunction or the $\since$-operator also
lead to undecidability, as we shall see shortly. 
Let us first fix the language with extended guarded quantifiers.
The syntax of quantified formulae is as follows:
$$
\forall \vec x : \psi(\vec x).\ \varphi(\vec x)
\qquad
\exists \vec x : \psi(\vec x).\ \varphi(\vec x).
$$
Here the formula $\psi(\vec x)$ is a guard, and $\vec x$ are
its only free variables. The semantics of the quantifiers are a straightforward
extension of that of $\ourLTL$, i.e., 
$$
\begin{array}{l}
   (h,i)  \models   \forall (x_1,\ldots,x_n): \psi(x_1,\ldots,x_n).\ \varphi 
\text{ iff } \\
\qquad   \text{ for all }
    c_1,\ldots,c_n, \text{ if } (h,i) \models \psi(c_1,\ldots,c_n)  
         \text{ then } (h,i) \models \varphi[x_1 := c_1, \ldots, x_n := c_n].
\end{array}
$$

Now consider a guarded quantifier that allows unrestricted uses of disjunction. 
Suppose $\varphi(\vec x)$, where $\vec x$ range over integers, is a formula encoding some
general Diophantine equation. Let $\psi(\vec x, y)$ be 
a guard formula $p(\vec x) \lor q(y)$, for some predicate $p$ and $q$ of appropriate
types. Then satisfiability of the entailment
$$
\{q(0) \} \models \exists (\vec x, y) : \psi(\vec x, y).\ \varphi(\vec x)
$$
is equivalent to the validity of the first-order formula
$
\exists \vec x.\ \varphi(\vec x),
$
which states the solvability of the Diophantine equations in $\varphi(\vec x).$
This means that the model checking problem for $\ourLTL$ with unrestricted disjunctive
guards is undecidable.
The cause of this undecidability is that satisfiability 
of the guard, relative to the history, is independent of the variables $\vec x.$
Similar observations can be made regarding the unrestricted uses of the ``since''
operator, e.g., if we replace the guard $\psi(\vec x, y)$ with
$p(\vec x) \since q(y)$, we get the same undecidability result. 

Another restriction that needs to be imposed on guarded quantifiers concerns
the use of function symbols: their uses easily lead to a violation of CWA, and
again, undecidability of model checking. For instance, in checking 
$$
\{p(0) \} \models \forall (x,y) : p(x + y).\ \varphi(x,y)
$$
we have to consider infinitely many combinations of $x$ and $y$ such
that $x + y = 0$.

Based on the above considerations, we design the following guarded extensions
to the quantifiers of $\ourLTL.$ The language of guards are defined 
as follows. {\em Simple guards} are formulae generated by the following
grammar:
$$
\gamma ::= p(\vec u) \mid \gamma \land \gamma \mid \always \gamma 
\mid \sometime \gamma 
$$
Here the list $\vec u$ is a list of variables and constants (no function
symbols allowed). We write $\gamma(\vec x)$ to denote a simple guard
whose only free variables are $\vec x.$
{\em Positive guards $G(\vec x)$ over variables $\vec x$} are formulae
whose only variables are $\vec x$, as generated by the following
grammar:
$$
G(\vec x) ::= \gamma(\vec x) \mid G(\vec x) \land G(\vec x)
\mid G(\vec x) \lor G(\vec x) 
\mid \always G(\vec x) \mid \sometime G(\vec x)
\mid G(\vec x) \since G(\vec x).
$$
We denote with $\ourLTLg$ the language obtained by extending $\ourLTL$ with
positive guards. We show that the model checking problem for $\ourLTLg$
is decidable. The key lemma to this 
is the finiteness of
the set of ``solutions'' for a guard formula. 

\begin{defn}
Let $G(\vec x)$ be a positive guard and let $h$ be a history.
The {\em guard instantiation problem}, written $(h, G(\vec x))$,
is the problem of finding a list $\vec u$ of constants such that
$h \models G(\vec u)$ holds. 
Such a list is called a {\em solution} of the guard instantiation
problem.
\end{defn}

\begin{lem}
\label{lm:guard}
Let $G(\vec x)$ be a positive guard over variables $\vec x$ and
let $h$ be a history. Then the set of solutions for the problem
$(h, G(\vec x))$ is finite. Moreover, every solution uses
only constants that appear in $h.$
\end{lem}
\vspace{-1em}
\begin{pf}
By induction on the size of $G(\vec x)$ and by definition of the forcing
relation.
\qed
\end{pf}

\vspace{-1em}

\begin{thm}
  Let $\varphi$ be a $\ourLTLg$ formula and $h$ a history.  
  The model checking problem $h \models \varphi$ is decidable.
\end{thm}
\vspace{-1em}
\begin{pf}
The proof follows the same structure as the decidability proof for
$\ourLTL$, using Lemma~\ref{lm:guard} for the quantifier cases.
\qed
\vspace{-1em}
\end{pf}


\vspace{-1em}

\section{Conclusions and related work}
\vspace{-0.5em}
\label{sec:conc}

We have presented a formal language for expressing history-based
access control policies based on the pure past
fragment of linear temporal logic, 
extended to allow certain guarded quantifiers and arbitrary computable
functions and relations. As our examples show, these extensions
allow us to write complex policies concisely,
while retaining decidability of
model checking.
Adding 
a counting quantifier 
allows us to express some statistical ``meta'' properties in policies.
We also consider the monitoring problem in the presence of
unobservable or unknown action parameters. 
We believe this is
the first formulation of the problem in the context of monitoring.

There is much previous work in the related area of history-based
access control
\cite{edjlali98ccs,havelund02tacas,fong04sp,bartoletti05fossacs,krukow05ccs,bauer:leucker:schallhart:FSTTCS06}.
As mentioned in the introduction, our transaction-based
approach to defining policies separates us from the more
traditional trace-based approaches in program execution monitoring.
Our work is closely related to Krukow, et al.\
\cite{krukow05ccs,krukow08jcs}, but there are a few important
differences. Their definition of sessions allows events to be
partially ordered using 
{\em event structures} \cite{winskel95handbook} whereas our notion of a session
as a set with no structure is simpler. For the application domains we
are interested in, we see no need for sessions to have extra structure
built into their semantics since such relations between events can be
explicitly encoded in our set up using first-order quantifiers and a
rich term language 
allowing
extra parameters, interpreted functions, timestamps and
arithmetic. In the first-order case, they forbid
multiple
occurrences of the same event in a session; roughly, their histories
in this case correspond to our
trace-like histories (see
Section~\ref{sec:modelchecking}).  
Their language does not allow arbitrary computable functions and
relations, since as we have seen, allowing these features in the
presence of quantifiers can easily lead to undecidability of model
checking. Our policy language is thus more expressive than theirs in
describing quantitative properties of histories as we have also seen
in some examples.

Although we have a prototypic implementation for checking histories
against policies of our language (cf.\ Section~\ref{sec:modelchecking}),
we plan to address further implementation related issues like
generating
more \emph{efficient} monitors that operate online
in the sense of \cite{havelund02tacas} for past-time LTL.
That is, when a
given trace is extended by a new session, an efficient monitor makes a
decision by merely
processing the extension rather than the previous history as well as the
extension. 
Another interesting
problem
is how to reason about policies, whereby we can tell that a policy is
subsumed by another, or when it is in conflict with another. This
requires finding a proof system for our 
logic which is sound and complete for our particular models (as finite
histories).


\vspace{-1em}


{\footnotesize\bibliographystyle{abbrv}
\bibliography{longstrings,bibliography}}

\setlength{\parindent}{0pt}
\setlength{\parskip}{0.0\baselineskip}

\appendix

\section{Detailed proofs}
\label{app:proofs}

In the following, given a history $h$,
we shall denote with $s(h)$ the {\em size} of $h$, i.e., the number
of symbols occuring in $h$.

\begin{lem}
The judgement $\mynode{h, i, \varphi}{\sat}$ is derivable if and only if
$(h,i) \models \varphi$. Similarly, the judgement $\mynode{ h, i, \varphi}{\unsat}$
is derivable if and only if $(h,i) \not \models \varphi.$
\end{lem}
\begin{pf}
  Straightforward by induction on the derivation tree of the evaluation
  judgements and the inductive definition of the semantic judgement $(h,i) \models
  \varphi.$ \qed
\end{pf}

\textbf{Theorem~\ref{thm:modelchecking} ~ }  
  Let $\varphi$ be a $\ourLTL$ formula and let $h$ be a history.  
  If the interpreted functions and relations in $\varphi$ are in \textsc{pspace}, then 
  the problem of deciding whether or not $h \models \varphi$
  holds is \textsc{pspace}-complete. 
\begin{pf}
  To show membership in \textsc{pspace}, we use Lemma~\ref{lm:eval} and show that
  checking the derivability of $\mynode{h, |h|,\varphi}{v}$, where $v$
  is either $\sat$ or $\unsat$, can be done in \textsc{pspace}.  Note that the
  transformation rules, reading them bottom-up, decrease the size of
  either the index $|h|$ or the size of $\varphi$, hence applying these
  transformations to the original judgement always terminates.  Moreover,
  the depth of any derivation tree for $\mynode{h,i,\varphi}{v}$ is
  bounded by $|h| + |\varphi|$. Note also that although the size of the
  derivation tree is exponential, one needs to check only one branch at
  a time. Therefore, to calculate the space requirement, we only need to
  calculate the space requirement for each node, multiplied by the
  maximum depth of the derivation tree.  At each node, we need to store
  the information about the child nodes that have not been visited, plus
  the values that have been computed for the child nodes that have been
  visited, which is a list of boolean values.  Notice that the branching
  factor of each rule, except $\forall$, is at most 3, and
  for $\forall$, it is at most $s(h).$ Therefore the branching factor
  of the rules is bounded by $s(h) + 3$, which means that the number of
  visited and not-yet-visited child nodes are also bounded by $s(h) + 3.$
  Each not-yet-visited child node takes up at most $s(h) + |\varphi|$ 
  space, since we need to store the history $h$ and an immediate subformulae
  of $\varphi$, and we need only to store the boolean value computed from
  each visited child node, which takes up a constant value (true or false), 
  say $b$. 
  Hence the space requirement for this model checking problem is at most
  $$
  (|h| + |\varphi|) \times (s(h) + 3) \times (s(h) + |\varphi| + b)
  $$
  which is polynomial in the size of $h$ and $\varphi$.

  To show \textsc{pspace}-hardness, we reduce the problem of checking
  satisfiability of quantified boolean formula (QBF), which is known to
  be \textsc{pspace} complete, to our model checking problem.  Let $F \equiv Q_1 x_1.\
  Q_2 x_2.\ \ldots Q_nx_n.\ E(x_1, x_2, \ldots, x_n)$ be a well-formed
  quantified Boolean formula (in prenex normal form), where $E$ is a
  Boolean expression involving variables $x_1,x_2,\ldots,x_n$, and $Q_i
  \in \{ \forall, \exists \}$.  The QBF problem then asks if $F$
  evaluates to $\top$ (cf.\ \cite{524279}).  Notice $F$ always evaluates
  to $\top$ or $\bot$ since there are no free variables in $F$.
  Let $F$ be given as above, then we construct in polynomial time a
  \ourLTL formula
  $$
  \varphi \equiv Q_1x_1:p_1.\ Q_2x_2:p_2.\ \ldots Q_nx_n:p_n.\
  E(true(x_1), true(x_2), \ldots, true(x_n)),
  $$
  and a history $h = \big\{ \{ p_1(0), p_1(1), p_2(0), p_2(1), \ldots, p_n(0),p_n(1),
  true(1) \} \big\}$,
  where $E$ uses the same Boolean connectives as in $F$.
  %
  It is then easy
  to see that $F$ evaluates to $\top$ if and only if $h \models
  \varphi$.
  \qed
\end{pf}


\textbf{Theorem~\ref{thm:fixed} ~ }  
  The decision problem for $h \models \varphi$, where $\varphi$ is
  fixed, is solvable in polynomial time in the size of $h$.
\begin{pf}
  Let the closure $cl(\varphi)$ of $\varphi$ contain all $m$
  subformulae of $\varphi$, i.e., $|cl(\varphi)| = m$, where $m$ is
  constant as $\varphi$ is fixed. For example, if $\varphi \equiv \Gp
  \forall x:p. \exists y:q.\ x \geq y$, then $cl(\varphi) = \{
  (\varphi), (\forall x:p. \exists y:q.\ x \geq y), (\exists y:q.\ x
  \geq y), (x \geq y) \}$.
  Then, to evaluate $h \models \varphi$, we first build a tree
  structure, similar to a syntax-tree, whose nodes correspond to the
  subformulae of $\varphi$ and whose root node corresponds to $\varphi$.
  We attach to each node, in a bottom-up manner, a truth table containing 
  the truth value of the subformula at the node, for all $|h|$ sessions 
  of the history and under all possible substitutions for the
  (free) variables occurring in the subformula.  Therefore, each table has $|h|$
  rows (one for each session), and at most $s(h)^m$ columns (since there
  are at most $m$ variables and each variable can range over at most
  $s(h)$ values). 
  We fill this table bottom up, from the first session and from 
  the atomic subformulae. The base case with atomic subformulae are easy;
  we need either to evaluate the relation symbols (in case it is an interpreted
  atomic formula) or perform a look up in the history. In both cases, it takes
  at most polynomial time. 
  By inspection of the semantics of $\ourLTL$, it is clear 
  that the truth value of a non-atomic formula depends on the truth value of its
  immediate subformulae, or on the same formula but at an earlier session.
  Thus to calculate the truth value of a non-leaf node at session $i$ and under
  a substitution $\sigma$, it is enough to perform table look up on 
  its immediate child nodes, or on earlier entries in the same node. 
  This requires at most linear time in the size of $h$. 
  Therefore, to fill up a truth table at each node, we need at most polynomial
  time. Since there are $m$ tables, the whole procedure takes at most polynomial
  time in the size of $h$.
  \qed
\end{pf}


\textbf{Theorem~\ref{thm:tracelike} ~ }  
  Let $\varphi$ be a $\ourLTL$ formula and $h$ be a trace-like history.
  The problem of deciding whether or not $h \models \varphi$ holds is
  \textsc{pspace}-complete in the size of $\varphi$.
\begin{pf}
  It is sufficient to show \textsc{pspace}-hardness. As in the proof of
  Theorem~\ref{thm:modelchecking}, we will map in polynomial time the
  \textsc{pspace}-complete QBF-problem to the given one, and show that
  the answer to the QBF-problem is $\top$ if and only if $h \models
  \varphi$ holds for carefully constructed $h$ and $\varphi$.

  Let $F \equiv Q_1 x_1. \ldots Q_nx_n.\ E(x_1, \ldots, x_n)$ be a
  QBF-problem defined as above.  Then, we construct a formula using the
  same connectives as in $E$,
  $$
  \varphi \equiv T_1Q_1x_1:p_1. \ldots T_nQ_nx_n:p_n.\ E(true(x_1),
  \ldots, true(x_n)),
  $$
  where $T_i$ is a temporal operator, and we have $T_i = \Gp$ if $Q_i =
  \forall$, and $T_i = \Fp$ if $Q_i = \exists$.  Furthermore, we
  construct a history
  $$
  h = \big\{ \{ p_n(0), true(1) \} \{ p_n(1), true(1) \}, \ldots, \hfill
  \{ p_1(0), true(1) \}, \{ p_1(1), true(1) \} \big\},
  $$
  where we separate different truth values in different sessions to
  preserve the trace-like structure.  To still be able to select
  different truth values for different predicates, we use temporal
  operators instead.
  So, if $h \models \varphi$ holds, then the $\Gp$ operator ensures that
  both interpretations of the corresponding predicate evaluate $F$ to
  $\top$, whereas $\Fp$ ensures that one of the two possible
  interpretations of the corresponding predicate evaluate $F$ to $\top$.
  %
  %
  \qed
\end{pf}

\textbf{Theorem~\ref{thm:counting} ~ }  
Assuming that the interpreted functions and relations are in \textsc{pspace},
the model checking problem for $\ourLTL$ extended with the counting 
quantifier is \textsc{pspace}-complete.
\begin{pf}
We need only to show membership in \textsc{pspace}. 
The proof follows the same outline as the proof of Theorem~\ref{thm:modelchecking},
but with the evaluation rule extended to deal with the counting quantifier:
$$
\mbox{
\urule{$\countq$}
{$\mynode{h,1,\psi}{v_1}
\quad 
\cdots
\quad
\mynode{h,i,\psi}{v_i}
\quad
\mynode{h,i,\varphi(n)}{v}
$
}
{$\mynode{h,i, \countq x : \psi. \varphi(x)}{v}$}
}
$$
where $n = \Sigma_{j = 1}^i I(v_i)$ and $I$ is a function
defined by $I(\sat) = 1$ and $I(\unsat) = 0.$ The branching
factor of this rule is bounded by $s(h) + 1$. The rest of the proof
proceeds as in the proof of Theorem~\ref{thm:modelchecking}.
\qed
\end{pf}


To prove Theorem~\ref{thm:partial}, we consider a slightly more general problem
$(h,i) \vdash \psi$ where $\psi$ can contain free variables, provided
they occur in $h$. The potential satisfiability problem is generalised
straightforwardly, i.e., $(h,i) \vdash_g \psi$ iff there exists a
substitution $\sigma$ such that $dom(\sigma) = V(h)$ and $(h\sigma,i)
\vdash \psi\sigma.$ 
In the following, given a finite set of formula $S$, 
we shall write $\bigvee S$ to denote the disjunction of 
all the formulae in $S$. In the case where $S$ is empty, 
$\bigvee S$ denotes $\bot.$
Likewise, $\bigwedge S$ denotes the conjunction of the formulae in $S$ and
when $S$ is empty, it denotes $\top.$

\begin{lem}
\label{lm:constraints}
For every $h$, $i$, and $\psi$, there exists a constraint formula $C$
such that $(h,i) \vdash_g \psi$ if and only if $C$ is satisfiable.
\end{lem}
\begin{pf}
  We construct $C$ by induction on $\psi$ and $i$.  If $i < 1$ or $i >
  |h|$ then $C = \bot.$ Obviously, $C$ is satisfiable iff $(h,i)
  \vdash_g \psi$.  We show some of the remaining cases here (the other
  cases are straightforward):
  \begin{enumerate}
  \item If $\psi$ is either $t_1 = t_2$, $t_1 \leq t_2$ or $t_1 \geq
    t_2$ then $C = \psi$.
  \item Suppose $\psi = p(t_1,\ldots,t_n).$ Then
    $$
    C = \bigvee \{u_1 = t_1 \land \cdots \land u_n = t_n \mid
    p(u_1,\ldots,u_n) \in h_i \}.
    $$




\item Suppose $\psi = \psi_1 \since \psi_2.$ By induction hypothesis (on
  the size of $\psi$) we have
  \begin{itemize}
  \item[(i)] $C_1$ such that $(h,i) \vdash_g \psi_1$ iff $C_1$ is
    satisfiable, and
  \item[(ii)] $C_2$ such that $(h,i) \vdash_g \psi_2$ iff $C_2$ is
    satisfiable.
  \end{itemize}
  If $i = 1$ then let $C = C_2$.  Otherwise, $i > 1$ and by induction
  hypothesis, we have $C_3$ such that $(h,i-1) \vdash_g \psi_1 \since
  \psi_2$ iff $C_3$ is satisfiable.  In this case, let $C = C_2 \lor
  (C_1 \land C_3).$

\item Suppose $\psi = \forall (x_1,\ldots,x_n) :
  p. \psi'(x_1,\ldots,x_n).$ By induction hypothesis, for each tuple
  $\vec u = (u_1,\ldots,u_n)$, we have a $C_{\vec u}$ such that $(h,i)
  \vdash_g \psi'(u_1,\ldots,u_n)$ iff $C_{\vec u}$ is satisfiable.
  Define $C$ as follows:
$$
C = \bigwedge \{C_{\vec u} \mid p(u_1,\ldots,u_n) \in h_i \}.
$$
\end{enumerate}
By inspecting the clauses of the forcing relation and the definition of
$(h,i) \vdash_g \psi$, it is straightforward to check that in each case
above $C$ is satisfiable if and only if $(h,i) \vdash_g \psi.$
\qed
\end{pf}


\end{document}